\documentclass[pdflatex,sn-mathphys-num]{sn-jnl}

\usepackage{graphicx}%
\usepackage{multirow}%
\usepackage{amsmath,amssymb,amsfonts}%
\usepackage{amsthm}%
\usepackage{mathrsfs}%
\usepackage[title]{appendix}%
\usepackage{xcolor}%
\usepackage{textcomp}%
\usepackage{manyfoot}%
\usepackage{booktabs}%
\usepackage{makecell}%
\usepackage{algorithm}%
\usepackage{algorithmicx}%
\usepackage{algpseudocode}%
\usepackage{listings}%
\usepackage{subcaption}%

\theoremstyle{thmstyleone}%
\theoremstyle{thmstyletwo}%

\theoremstyle{thmstylethree}%

\begin{document}

\title[Article Title]{Privacy-Preserving Chest X-ray Report Generation via Multimodal Federated Learning with ViT and GPT-2}


\author*[1]{\fnm{Md. Zahid} \sur{Hossain}}\email{zahid.cse@aust.edu}
\equalcont{These authors contributed equally to this work.}

\author[1]{\fnm{Mustofa} \sur{Ahmed}}\email{mustofahmed24@gmail.com}
\equalcont{These authors contributed equally to this work.}

\author[2]{\fnm{Most. Sharmin Sultana} \sur{Samu}}\email{sharminsamu130@gmail.com}
\equalcont{These authors contributed equally to this work.}

\author[1]{\fnm{Md. Rakibul} \sur{Islam}}\email{rakib.aust41@gmail.com}
\equalcont{These authors contributed equally to this work.}

\affil[1]{\orgdiv{Department of Computer Science and Engineering}, \orgname{Ahsanullah University of Science and Technology}, \orgaddress{\city{Dhaka}, \postcode{1208}, \country{Bangladesh}}}

\affil[2]{\orgdiv{Department of Computer Science and Engineering}, \orgname{BRAC University}, \orgaddress{\city{Dhaka}, \postcode{1212}, \country{Bangladesh}}}


\abstract{The automated generation of radiology reports from chest X-ray images holds significant promise in enhancing diagnostic workflows while preserving patient privacy. Traditional centralized approaches often require sensitive data transfer, posing privacy concerns. To address this, the study proposes a Multimodal Federated Learning framework for chest X-ray report generation using the IU-Xray dataset. The system utilizes a Vision Transformer (ViT) as the encoder and GPT-2 as the report generator, enabling decentralized training without sharing raw data. Three Federated Learning (FL) aggregation strategies—FedAvg, Krum Aggregation and a novel Loss-aware Federated Averaging (L-FedAvg)—were evaluated. Among these, Krum Aggregation demonstrated superior performance across lexical and semantic evaluation metrics such as ROUGE, BLEU, BERTScore and RaTEScore. The results show that FL can match or surpass centralized models in generating clinically relevant and semantically rich radiology reports. This lightweight and privacy-preserving framework paves the way for collaborative medical AI development without compromising data confidentiality.}




\maketitle

\section{Introduction}\label{sec1}

Radiology reports are essential diagnostic tools in modern medicine. They play a critical role in patient care and treatment planning. Among various radiological techniques, X-rays are the most commonly used imaging modality in clinical practice. Chest X-rays (CXRs) are particularly popular due to their speed, cost-effectiveness and diagnostic value. CXRs help identify abnormalities in the lungs, heart, ribs and surrounding structures. Common findings include pneumonia, pneumothorax, cardiomegaly, broken ribs and consolidation etc.

Traditionally, the interpretation of X-ray images requires the expertise of trained radiologists. However, with the advent of deep learning (DL), computer-aided diagnosis has emerged as a powerful tool to support and enhance clinical decision-making. Deep learning models have demonstrated remarkable success in analyzing medical images and detecting diseases with accuracy comparable to that of human experts \cite{LITJENS201760}. In parallel, advances in generative artificial intelligence (AI) have enabled the development of models capable of automatically generating detailed radiology reports from X-ray images \cite{chen2022cross} \cite{jing2017automatic} \cite{li2019knowledge}.

Despite significant progress in report generation, most existing approaches rely on centralized model training. This setup requires transferring sensitive patient data from hospital systems to external research facilities, which raises ethical and legal concerns regarding patient privacy and data security. As a result, many healthcare institutions are reluctant to share data, limiting the development and deployment of robust AI systems \cite{Murdoch2021}.

Federated Learning (FL) \cite{mcmahan2017communication} offers a promising solution to these challenges. It is a distributed machine learning paradigm in which multiple clients (e.g., hospitals) collaboratively train a global model without exchanging raw data. Instead, only model updates are shared, preserving data locality and aligning with data protection regulations such as HIPAA and GDPR \cite{kairouz2021advancesopenproblemsfederated}. While FL has been extensively applied to disease classification from chest X-rays \cite{chakravarty2021federated} \cite{banerjee2020multi}  \cite{sheller2020federated} \cite{slazyk2022cxr} \cite{ziegler2022defending}, its application to radiology report generation remains largely unexplored. This is due to several inherent challenges, including handling non-independent and identically distributed (non-IID) data, communication overhead, model heterogeneity and the need for clinically accurate and interpretable reports.

Motivated by these privacy concerns and the growing capabilities of generative models, we explore the application of federated learning in radiology report generation from chest X-ray images. In this work, we propose a novel federated learning framework for radiology report generation using chest X-ray data. Unlike conventional approaches, our method does not require high-end distributed systems. Instead, it leverages simple, accessible tools like Google Drive and Firebase for parameter exchange and data handling. To the best of our knowledge, no prior work has demonstrated such a lightweight and replicable implementation of FL for this task.
\\\\Our key contributions are as follows:
\begin{itemize}
\item  We introduce a novel, practical implementation of federated learning for radiology report generation using chest X-ray images.
\item We evaluate the quality of generated reports using standard natural language generation metrics, including BLEU \cite{papineni2002bleu}, ROUGE \cite{lin2004rouge}, BERTScore \cite{zhang2019bertscore} and RaTEScore \cite{zhao-etal-2024-ratescore}.

\item We show that our federated approach can outperform baseline centrally trained models, demonstrating the feasibility of secure, decentralized training in sensitive medical applications.
\item Our method aims to reduce the workload of clinicians by automating the report-writing process, while also enabling healthcare institutions to collaboratively train high-quality models without compromising patient data privacy.
\end{itemize}

This article is organized into several sections. Section \ref{sec2} provides a summary of related studies. The research approach is explained in Section \ref{sec3}. Section \ref{sec4} describes the experimental setup including dataset details. Section \ref{sec5} presents the research outcomes and compares the performance of various federated learning aggregation techniques. Finally, Section \ref{sec6} discusses the study's limitations and suggests directions for future research.

\section{Related Work}\label{sec2}

Federated Learning has emerged as a promising approach for medical imaging applications particularly in chest radiograph classification by enabling privacy-preserving model training across multiple institutions. \cite{chakravarty2021federated} employs a CNN-GNN framework to address data heterogeneity and co-morbidity dependencies using CheXpert data partitioned across five sites. The model modifies Federated Averaging by training site-specific GNNs leading to a 1.74\% performance improvement by achieving an average AUC of 0.79. However, model generalizability remains a challenge due to site-specific data distributions and future work should integrate clinical priors. \cite{banerjee2020multi} focuses on pneumonia classification using FL with transfer learning on ResNet18, ResNet50, DenseNet121 and MobileNetV2. The models achieve 98.3\% accuracy for pneumonia detection and 87.3\% for bacterial-viral differentiation with Momentum SGD outperforming adaptive optimizers. Future improvements include extending classification to more diseases and optimizing hybrid FL frameworks. \cite{sheller2020federated} compares FL with institutional incremental learning (IIL) and cyclic IIL (CIIL) using radiographic imaging from ten institutions. It demonstrates FL’s ability to achieve 99\% of centralized model performance. The study highlights FL’s effectiveness but notes biases from institutional data variations and insufficient synchronization in some data types. Future directions include hyper-parameter tuning and addressing institutional biases.

\cite{slazyk2022cxr} focuses on FL for chest X-ray analysis using the RSNA 2018 dataset. It employs UNet++ with EfficientNet-B4 for segmentation, ResNet50 and DenseNet121 for classification. The study finds that FL improves generalizability. ResNet50 achieved 0.757 accuracy but highlights challenges in optimizing client selection and training epochs. \cite{ziegler2022defending} integrates differential privacy into FL to counter reconstruction attacks in chest X-ray classification using the Mendeley and CheXpert datasets. While differential privacy reduces data leakage, privacy risks remain. The study suggests refining privacy budget optimization. \cite{ho2022fedsgdcovid} extends FL to COVID-19 detection by combining chest X-ray images with symptom data. It implements CNN with spatial pyramid pooling and Differential Privacy Stochastic Gradient Descent (DP-SGD) but observes accuracy degradation in non-IID datasets. It emphasizes the need for robustness improvements. \cite{linardos2022federated} applies FL to cardiovascular disease diagnostics utilizing a 3D-CNN model pretrained on action recognition and the FL-EV voting approach across four medical centers. The study finds that FL-EV enhances model performance particularly in larger centers but is constrained by the small dataset size. \cite{adnan2022federated} examines FL for histopathology image analysis using the TCGA dataset integrating FedAvg with differential privacy techniques like Rényi Differential Privacy Accountant. The study confirms that FL achieves comparable performance to centralized training while addressing privacy concerns especially in non-IID scenarios. A common limitation across these studies is performance degradation in non-IID settings and challenges in privacy-preserving mechanisms. Future research should refine FL parameters, improve data distribution strategies, optimize privacy budgets and explore additional security techniques to enhance model robustness and generalizability in real-world medical applications.

\cite{tayebi2023enhancing} evaluates CNNs and transformer-based architectures using a large dataset of 610,000 chest radiographs from five institutions. It highlights FL’s role in improving off-domain performance emphasizing the impact of data diversity. In contrast, \cite{chowdari2023federated} employs the DenseNet-121 architecture on publicly available datasets (NIH, VinBigData and CheXpert). It demonstrates improved model generalizability with a novel aggregation scheme. SecureFed, introduced in \cite{makkar2023securefed} enhances lung abnormality analysis through secure aggregation. It outperforms existing frameworks like FedAvg, FedMGDA+, FedRAD in robustness and fairness with evaluation on a COVID-19 dataset. Meanwhile, \cite{sohan2023systematic} presents a systematic literature review of FL applications in medical imaging focusing on privacy preservation and performance evaluation. It underscores FL’s effectiveness in securing medical data. \cite{jindal2023pushing} integrates FL with CNNs, specifically VGG-16, to diagnose lung diseases by employing focal loss to address data imbalance. It achieved high accuracy (88.43\%-96.69\%) across different clients. A common limitation across these studies is the challenge of handling data heterogeneity. Some works \cite{tayebi2023enhancing} \cite{makkar2023securefed} \cite{jindal2023pushing} emphasize the need for validation on larger and more diverse datasets. Future research directions include integrating imaging with non-imaging features \cite{tayebi2023enhancing}, refining aggregation techniques \cite{chowdari2023federated}, exploring scalability \cite{makkar2023securefed}, conducting experimental validations \cite{sohan2023systematic} and extending FL applications to other medical domains \cite{jindal2023pushing}.

\cite{liu2023fedarc} introduces FedARC, a personalized FL method that integrates adaptive regularization and model-contrastive learning to improve tuberculosis (TB) diagnosis accuracy. It demonstrates superiority over FedAvg. \cite{zubair2023comparative} compares FL with centralized learning for lung disease detection. It utilizes an ensemble of deep learning models such as VGG19, DenseNet and Inception. It shows that FL can achieve comparable or better performance while preserving privacy. \cite{ullah2023scalable} presents a scalable FL framework that incorporates data augmentation techniques to address imbalanced datasets by achieving 98.14\% accuracy but facing challenges with client participation variability. \cite{malik2023dmfl_net} proposes DMFL\_Net which employs DenseNet-169 for feature extraction. It outperformed VGG models with an accuracy of 98.45\% yet struggled with model bias due to non-IID data. \cite{nazir2023federated} provides a survey of FL in medical image analysis emphasizing the integration of deep neural networks (DNNs) and discussing the potential of data augmentation and GANs to enhance model generalization while identifying security concerns and dataset biases as critical challenges. Across these studies, FL consistently demonstrates its potential to enhance disease diagnosis while maintaining data privacy. However, challenges such as data heterogeneity, model aggregation inefficiencies and dataset biases persist. Future research directions include refining aggregation strategies, optimizing communication efficiency, integrating multimodal data and improving model robustness across diverse medical settings.

\cite{chen2024medical} introduces a multimodal FL framework for medical report generation integrating deep learning and the federal average algorithm which enhances report accuracy while preserving patient privacy. Its limitations include scalability and dataset validation. \cite{khan2024federated} applies FL with ResNet-50 for disease prediction from chest X-rays. It improves detection accuracy by 2\% but is limited by dataset diversity. \cite{babar2024federated} combines FL with active and transfer learning (FAL-TL) for lung cancer diagnosis. It achieves remarkable accuracies of 99.20\% and 98.70\% on different datasets yet struggles with scalability and communication overhead. \cite{guan2024federated} provides a comprehensive survey of FL in medical imaging categorizing methods and highlighting computational and communication challenges while advocating for broader FL applications. Recent works have applied FL for privacy-preserving health monitoring on smartphones. \cite{tabassum2023depression} used smartphone sensor data for depression detection but lacked multi-modal inputs. \cite{ahmed2025ondevicefederatedlearningsmartphones} employed FL on Reddit text data, but reliance on social media limits its clinical applicability. While both focus on sensor or text data, they highlight the growing interest in decentralized health AI. Building on this, we explore vision-language models for X-ray report generation using a ViT-GPT2 architecture within an FL framework.

In terms of models, the studies predominantly leverage Convolutional Neural Networks (CNNs) for tumor detection \cite{alruwais2025federated, kumari2025cnn}, emotion prediction \cite{raju2024enhancing} and brain tumor diagnosis \cite{onaizah2025fl, alalwan2024advancements}. Additionally, Siamese CNNs (SiCNN) \cite{onaizah2025fl, alanazi2025enhancing} and Generative Adversarial Networks (GANs) \cite{alalwan2024advancements} are used for improving privacy and data augmentation. Federated learning plays a crucial role in enabling collaborative model training without sharing raw data, which is central to preserving privacy across decentralized healthcare environments \cite{alruwais2025federated, kumari2025cnn, onaizah2025fl, alanazi2025enhancing, alalwan2024advancements, raju2024enhancing}. The datasets employed vary from pancreatic tumor datasets \cite{alruwais2025federated} and MRI scans \cite{onaizah2025fl, alanazi2025enhancing, alalwan2024advancements} to emotion prediction datasets like RAVDESS \cite{raju2024enhancing} and healthcare datasets from Saudi Arabia \cite{rathee2025improved} to highlight the varied applications of FL across different domains. Performance results show high accuracy rates, such as 99.82\% for brain tumor detection \cite{alalwan2024advancements} and 95.72\% for emotion prediction \cite{raju2024enhancing}, though the studies acknowledge limitations such as small, imbalanced datasets and issues with scalability and generalizability \cite{alruwais2025federated, alalwan2024advancements, raju2024enhancing, rathee2025improved}. In terms of future work, expanding dataset diversity, enhancing model robustness and testing the frameworks across varied real-world healthcare scenarios are commonly recommended \cite{alruwais2025federated, onaizah2025fl, alanazi2025enhancing, rathee2025improved}. Notably, the integration of emerging technologies such as 5G \cite{kumari2025cnn}, GANs \cite{alalwan2024advancements} and zero-shot learning \cite{rathee2025improved} further distinguishes these studies with future research aiming to refine these models for broader applicability in real-world healthcare settings. 

\cite{naz2024centralized} compares traditional deep learning (DL) models with FL for COVID-19 detection using chest X-ray images from the Radiography CXR dataset. While models like ResNet50 achieved an accuracy of 98\%, FL demonstrated slightly lower accuracy (3.56\% reduction) due to its handling of non-IID datasets but showed faster convergence and better performance with more clients. Limitations include the absence of a detailed discussion on communication costs and the need for model tuning. \cite{muthalakshmi2024federated} proposes a federated learning framework that incorporates secure techniques such as differential privacy and homomorphic encryption. It uses real-world medical imaging datasets and achieves an accuracy of 98.6\% with a focus on secure data sharing. Challenges such as communication overhead and model convergence are noted. Future research aims to enhance FL efficiency. \cite{adhikari2401secure} focuses on FL for COVID-19 diagnosis using the COVIDGR and CheXpert datasets employing DenseNet-121 and Grad-CAM for interpretability. While achieving good results, the study identifies issues with non-IID data distributions and calls for improvements in algorithmic robustness such as FedProx or Scaffold to enhance generalization. \cite{ram2024federated} investigates the use of federated learning with the NIH Chest X-ray dataset using a ResNet-34 model with secure aggregation and homomorphic encryption. The study reports a clinical-grade accuracy of 83\% but acknowledges the need for more clients and scalability in real-world settings. All above studies emphasize the potential of FL in maintaining patient privacy while achieving high diagnostic performance, yet all identify similar challenges regarding data distribution, communication costs and model generalization. Future work across these papers points to improvements in scalability, efficiency and robust handling of non-IID data.\\
\\
The following research gaps are identified through our extensive literature search:
\begin{itemize}
\item  Federated learning for chest X-ray report generation is underexplored with no clear implementation guides available.
\item FL models lack generalizability due to site-specific data and limited validation on diverse, large datasets.
\item Addressing data heterogeneity, non-IID distributions and institutional biases remains a challenge in FL.
\item Scalability, communication efficiency and convergence issues hinder FL implementations.
\item FL research lacks focus on hybrid frameworks, robust model tuning and combining imaging with non-imaging data.
\end{itemize}

\section{Methodology}\label{sec3}

\begin{figure}[h]
  \centering
  \includegraphics[width=1.0\linewidth]{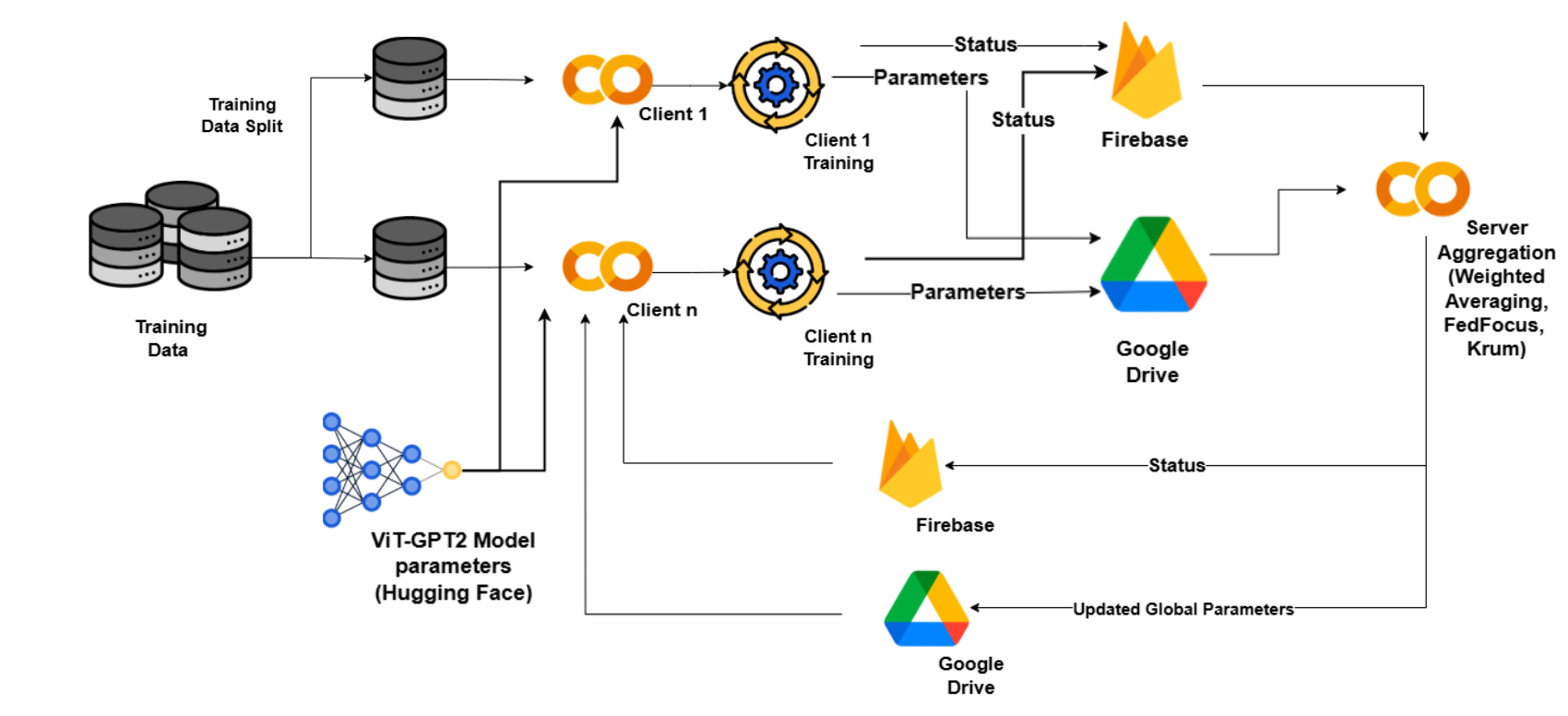}  
  \caption{Working Approach of our Federated Learning.}
  \label{fig:methodology}
\end{figure}

Our federated learning framework adheres to the workflow depicted in Figure~\ref{fig:methodology}. The process begins with image acquisition and preprocessing. Chest X-ray images are sourced from the publicly available IU-Xray dataset \cite{chen2020generating}, ensuring both diversity and task relevance. The dataset is partitioned among four clients, as summarized in Table~\ref{tab:client_data_distribution}.

\begin{table}[ht]
  \caption{Client Data Distribution}
  \label{tab:client_data_distribution}
  \centering
  \begin{tabular}{l c c}
    \toprule
    \textbf{Client} & \textbf{Training Data Size} & \textbf{Validation Data Size} \\
    \midrule
    Client 1 & 1655 & 237 \\
    Client 2 & 1241 & 178 \\
    Client 3 & 828 & 117 \\
    Client 4 & 414 & 60 \\
    \bottomrule
  \end{tabular}
\end{table}

\begin{table*}[ht]
  \caption{Client's Model Training Configuration}
  \label{tab:training_configuration}
  \centering
  \small
  \begin{tabular}{lccccc}
    \toprule
    \textbf{Model} & 
    \makecell{\textbf{Epochs} \\ \textbf{per Round}} & 
    \makecell{\textbf{Training} \\ \textbf{Batch Size}} & 
    \textbf{Optimizer} & 
    \makecell{\textbf{Learning} \\ \textbf{Rate}} & 
    \makecell{\textbf{Weight} \\ \textbf{Decay}} \\
    \midrule
    ViT B16+GPT2 & 3 & 8 & AdamW & 5e-5 & 0.01 \\
    \bottomrule
  \end{tabular}
\end{table*}
Each client is allocated a subset of the training and validation data, along with pretrained model parameters for the Vision Transformer (ViT)~\cite{dosovitskiy2020image} and GPT-2~\cite{radford2019language}. These pretrained models are retrieved from the Hugging Face library. Clients are simulated independently using Google Colab notebooks. Upon receiving the global pretrained parameters and corresponding data, each client initiates local training. The hyperparameters employed by each client are given in Table~\ref{tab:training_configuration}.

The ViT model is utilized for visual feature extraction. It segments each image into fixed-size patches, which are embedded into feature vectors and subsequently processed through self-attention layers to capture global contextual information. These visual features serve as input to the language generation module.

GPT-2 is employed for text generation, functioning as a decoder-only architecture. It generates descriptive text in an autoregressive manner, conditioned on the extracted visual features. A cross-attention mechanism is incorporated to effectively integrate visual representations into the language model, thereby ensuring semantic alignment between the input image and the generated text. The models are trained using paired image-report data to learn the mapping between visual content and textual descriptions.

As the system operates synchronously, all clients must complete their local training before progressing to the next federated round. Upon completion, each client uploads its locally updated model parameters to a shared Google Drive folder. Although Firebase Storage was initially considered, it was found to be less efficient for uploading large files, so Google Drive was used instead. Given that the models are initialized from large pretrained checkpoints, each parameter file ranges between 700 and 800 MB.

In addition to uploading model parameters, each client updates its training status in the Firebase Realtime Database. The central server, also implemented in a Colab notebook, continuously monitors the database to determine when all clients have completed their training. Once confirmation is received, the server performs model aggregation using one of three federated strategies: Krum Aggregation~\cite{blanchard2017machine}, Federated Averaging (FedAvg)~\cite{mcmahan2017communication}, or Loss-aware Federated Weighted Averaging, our novel approach.

Following aggregation, the updated global model parameters are uploaded to the shared Google Drive folder. The server also posts a status update in the Firebase Realtime Database, prompting clients to begin the next training round. Clients monitor the database for this update, download the aggregated global model and resume training. This process is iteratively repeated across multiple rounds.

The quality of the generated radiology reports is evaluated using a combination of lexical and semantic similarity metrics. ROUGE is employed to assess n-gram overlap and precision, while BERTScore is used to measure semantic similarity between generated and reference texts. Additionally, RateScore is used to evaluate the factual consistency and clinical relevance of the generated reports by comparing them to ground truth findings.

\subsection{Federated Learning Algorithms}

\subsubsection{Federated Weighted Averaging}

This is the vanilla federated learning algorithm. The federated learning algorithm was initially developed by a group of researchers from Google \cite{mcmahan2017communication}. This algorithm is written by the authors who first proposed federated learning. Here, each client contributes based on the proportion of the size of the dataset they have. The clients that have the largest data size during training contribute the highest to the global model. All the clients send their model parameters to the global model, the global model will aggregate the model parameters by averaging all the client parameters while assigning more weights to the clients that have higher dataset size. 

\begin{algorithm}[H]
\caption{Federated Averaging (FedAvg) Algorithm}
\begin{algorithmic}[1]
\Require Number of communication rounds $T$, number of clients $K$, fraction of clients $C$, local epochs $E$, learning rate $\eta$
\State Initialize global model parameters $\mathbf{w}_0$
\For{each round $t = 1, \dots, T$}
    \State Randomly select a subset of clients $\mathcal{S}_t \subseteq \{1, \dots, K\}$ with $|\mathcal{S}_t| = \max(C \cdot K, 1)$
    \For{each client $k \in \mathcal{S}_t$ \textbf{in parallel}}
        \State $\mathbf{w}_t^k \gets \mathbf{w}_t$ \Comment{Send global model to client $k$}
        \State $\mathbf{w}_t^k \gets \text{ClientUpdate}(k, \mathbf{w}_t^k)$
    \EndFor
    \State $\mathbf{w}_{t+1} \gets \frac{1}{|\mathcal{S}_t|} \sum_{k \in \mathcal{S}_t} \mathbf{w}_t^k$ \Comment{Aggregate updates}
\EndFor
\State \Return Final global model parameters $\mathbf{w}_T$

\Function{ClientUpdate}{$k, \mathbf{w}$}
    \State $\mathcal{B} \gets$ (split local data $\mathcal{D}_k$ into batches)
    \For{each local epoch $e = 1, \dots, E$}
        \For{each batch $b \in \mathcal{B}$}
            \State $\mathbf{w} \gets \mathbf{w} - \eta \nabla \ell(\mathbf{w}; b)$ \Comment{Gradient descent}
        \EndFor
    \EndFor
    \State \Return $\mathbf{w}$
\EndFunction

\end{algorithmic}
\end{algorithm}

\subsubsection{Krum Aggregation}

The Krum aggregation algorithm in federated learning is a robust method that can defend itself against adversarial attacks or clients \cite{blanchard2017machine}. This is an improved version of the vainlla federated learning algorithm. It works by selecting a client update (gradient or parameter vector) that is least affected by malicious updates. For all the clients, the Krum calculates or measures the distance to all the other clients and sums them up. It then considers the closest one, which is the value with least sum for aggregation. This approach improves the security by removing malicious clients. But this is computationally very expensive because each client needs to calculate distance to all other clients. So although this is a good algorithm when there is a chance of potential malicious clients but it comes at a cost. 

\begin{algorithm}[H]
\caption{Krum Aggregation Algorithm}
\begin{algorithmic}[1]
\Require Global model parameters $\mathbf{w}_t$, updates from $m$ clients $\{\mathbf{w}_t^1, \dots, \mathbf{w}_t^m\}$, number of clients $m$, number of Byzantine clients $f$
\Ensure Aggregated model parameters $\mathbf{w}_{t+1}$
\State Initialize an empty list $S$
\For{each client $k \in \{1, \dots, m\}$}
    \State Compute distances $d_{k,j} = \|\mathbf{w}_t^k - \mathbf{w}_t^j\|^2 \, \forall j \neq k$
    \State Sort distances $d_{k,j}$ in ascending order
    \State Compute the sum of the $m-f-2$ smallest distances: $D_k = \sum_{j=1}^{m-f-2} d_{k,j}$
    \State Append $(k, D_k)$ to $S$
\EndFor
\State Select the client $k^*$ with the smallest $D_k$ from $S$
\State $\mathbf{w}_{t+1} \gets \mathbf{w}_t^{k^*}$
\State \Return $\mathbf{w}_{t+1}$
\end{algorithmic}
\end{algorithm}

\subsubsection{Loss-aware Federated Weighted Averaging}

We propose Loss-aware Federated Weighted Averaging (L-FedAvg), an enhanced variant of the standard federated averaging algorithm. This robust federated learning approach prioritizes clients whose learning objectives are more closely aligned with the global model. Before aggregation, the server evaluates each client's validation loss in conjunction with their total number of training samples. Clients with lower validation loss are given higher priority by assigning them greater weights, while those with higher loss receive reduced weights. As a result, clients that are better aligned with the global objective have a greater influence on the global model update, improving convergence and overall performance. 
\begin{table*}[ht]
  \caption{FL Aggregation Hyperparameters}
  \label{tab:fl_aggregation_parameters}
  \centering
  \footnotesize
  \begin{tabular}{l c c p{6.5cm}}
    \toprule
   \textbf{Approach} & 
    \makecell{\textbf{Parameter} \\ \textbf{Name}} & 
    \textbf{Value} & 
    \textbf{Purpose} \\
    \midrule
    L-FedAvg & alpha & 0.5 & Controls weightage of validation loss and training data size \\
    Krum & Fault Tolerance & 1 & Tolerates up to 1 faulty client during aggregation \\
    \bottomrule
  \end{tabular}
\end{table*}\\
The hyperparameters used in different approaches shown in Table \ref{tab:fl_aggregation_parameters}.

\begin{algorithm}[H]
\caption{Loss-aware Federated Weighted Averaging}
\begin{algorithmic}[1]
\Require List of client model weights $\{\mathbf{w}_k\}_{k=1}^K$, JSON file with client data, weighting factor $\alpha \in [0,1]$
\Ensure Aggregated global model weights $\mathbf{w}_{\text{avg}}$
\State Extract data lengths $\{d_k\}$ and validation losses $\{l_k\}$ from JSON
\State Compute total data points $D = \sum_{k=1}^K d_k$
\For{each client $k = 1, \dots, K$}
    \State Compute data weight: $w_d^k = \frac{d_k}{D}$
    \State Compute loss weight: $w_l^k = \frac{1}{l_k}$ \Comment{If $l_k > 0$}
    \State Compute combined weight: $w_k = \alpha \cdot w_d^k + (1 - \alpha) \cdot w_l^k$
\EndFor
\State Normalize weights: $w_k \gets \frac{w_k}{\sum_{j=1}^K w_j}$ for all $k$
\State Initialize $\mathbf{w}_{\text{avg}} \gets 0$
\For{each parameter $\theta$ in the model}
    \State $\mathbf{w}_{\text{avg}}[\theta] \gets \sum_{k=1}^K w_k \cdot \mathbf{w}_k[\theta]$
\EndFor
\State \Return $\mathbf{w}_{\text{avg}}$
\end{algorithmic}
\end{algorithm}

\subsection{Pretrained Models}

\subsubsection{Vision Transformer (ViT)}

The Vision Transformer (ViT) is a model that applies transformer architecture, originally designed for natural language processing tasks, to computer vision. Instead of processing entire images as a grid of pixels, ViT divides images into smaller patches, treats each patch as a sequence and processes them using transformer layers \cite{dosovitskiy2020image}. This approach enables ViT to model relationships between patches, capturing global context efficiently. ViT has shown competitive performance compared to traditional convolutional neural networks (CNNs), especially when trained on large datasets.

\subsubsection{GPT-2 (Generative Pre-trained Transformer 2)}

GPT-2 (Generative Pre-trained Transformer 2) is a language model designed for generating human-like text \cite{radford2019language}. It is based on the transformer architecture and trained on a large corpus of text from the internet. GPT-2 uses unsupervised learning to predict the next word in a sentence, allowing it to understand context and generate coherent and contextually relevant text. The model can perform a wide range of natural language processing tasks, such as text completion, summarization and translation, without task-specific training.

\section{Experimental Setup} \label{sec4}

The experiments were conducted using NVIDIA T4 GPUs provided by Google Colab. Standard deep learning libraries such as PyTorch and Hugging Face Transformers were used. All the clients and the server had the same setup. Model parameters were stored securely on Google Drive. Firebase Database was used for seamless communication by handling status updates between the server and clients.

\subsection{Dataset}
\begin{figure}[h]
  \centering
  \includegraphics[width=0.85\textwidth]{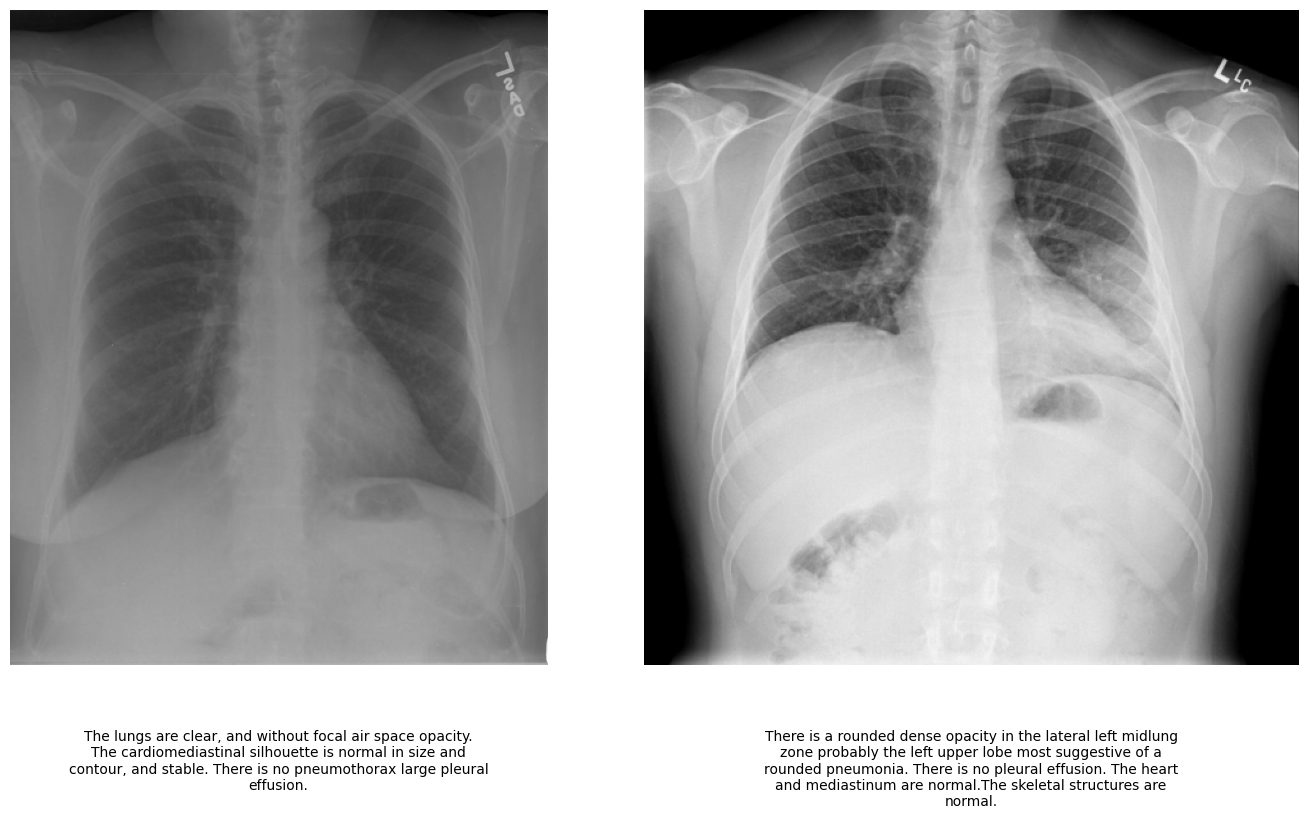}
  \caption{Sample X-ray images and corresponding findings in form of report from the IU-Xray dataset. This report is being treated as the ground truth. }
  \label{fig2}
\end{figure}

We have used the IU-Xray \cite{chen2020generating} dataset which is a publicly available collection of radiographic images paired with their corresponding radiology reports. This dataset is widely used for research in medical imaging. The dataset comprises a total of 5,910 chest X-ray images along with their associated findings in the form of radiology reports. Each image in the dataset is accompanied by a detailed textual description that provides diagnostic insights. Figure~\ref{fig2} presents two sample cases from the dataset: one depicting a normal chest X-ray and the other showing an abnormal case, along with their corresponding reports.

The dataset is organized into predefined splits for training, testing and validation. Training set, test set and validation set contains 4138, 1180 and 592 images and their corresponding reports respectively.

\begin{table}[ht]
  \caption{Statistics of Report Length by Split}
  \label{tab:Report Length by Split}
  \centering
  \begin{tabular}{l c c c c c c c}
    \toprule
    \textbf{Split} & \textbf{Count} & \textbf{Mean} & \textbf{STD} & \textbf{Min} & \textbf{Max} & \textbf{25\%} & \textbf{50\%} \\
    \midrule
    Train & 4138.0 & 31.765 & 14.206 & 7.0 & 149.0 & 22.0 & 29.0 \\
    Test & 1180.0 & 28.219 & 13.181 & 8.0 & 93.0 & 19.0 & 25.0 \\
    Validation & 592.0 & 31.128 & 13.812 & 8.0 & 83.0 & 21.0 & 30.0 \\
    \bottomrule
  \end{tabular}
\end{table}

\begin{figure}[h]
  \centering
  \includegraphics[width=0.7\textwidth]{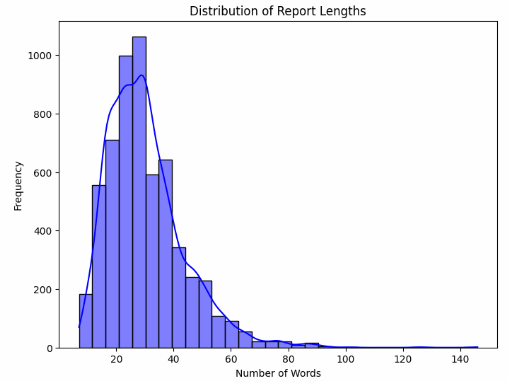}
  \caption{Distribution of report length in number of words}
  \label{fig3}
\end{figure}

\begin{figure}[H]
  \centering
  \includegraphics[width=0.7\textwidth]{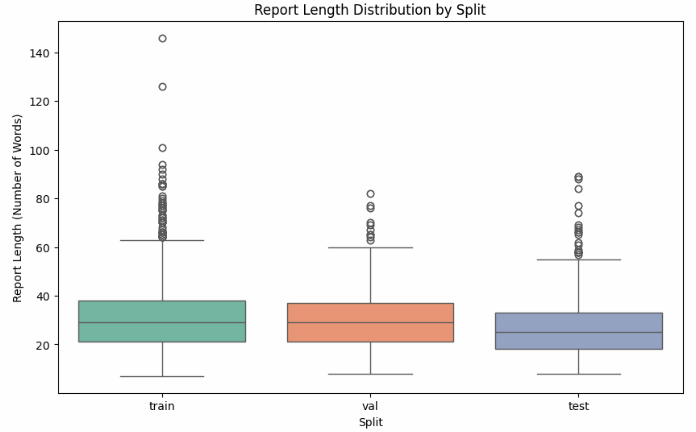}
  \caption{Report length distribution in train, test and validation split}
  \label{fig4}
\end{figure}

Before conducting our experiments, we performed an initial exploration of the dataset to better understand the characteristics of the radiology reports. Table~\ref{tab:Report Length by Split} summarizes key statistics of report lengths across the training, validation and test splits, including mean, standard deviation and percentiles. Figure~\ref{fig3} shows the overall distribution of report lengths in terms of word count and their corresponding frequencies. Additionally, a boxplot is presented in Figure~\ref{fig4} to illustrate the variation in report length across the train, validation and test sets.

\section{Result Analysis}\label{sec5} 
Table \ref{tab:Model Evaluation-1} shows the performance of various approaches. The performance was evaluated using multiple different n-gram based metrics including ROUGE and BLEU. Among all the approaches, Krum Aggregation achieved the highest scores in most of the metrics. This approach obtained the best ROUGE-1 F1 score of 0.306, ROUGE-2 F1 score of 0.1411 and ROUGE-3 F1 score of 0.0726. This approach obtained a BLEU score of 0.0395 and ROUGE-L score of 0.2066.

\begin{table*}[ht]
 \centering
  \caption{Evaluation with N-gram Based Metrics}
  \label{tab:Model Evaluation-1}
  \footnotesize
  \begin{tabular}{l c c c c c c}
    \toprule
    \textbf{Approach} & \makecell{\textbf{ROUGE1} \\ \textbf{F1}} & \makecell{\textbf{ROUGE2} \\ \textbf{F1}} & \makecell{\textbf{ROUGE3} \\ \textbf{F1}} & \makecell{\textbf{ROUGE4} \\ \textbf{F1}} & \makecell{\textbf{ROUGEL} \\ \textbf{F1}} & \textbf{BLEU} \\
    \midrule
    FedAvg &  0.2928 & 0.1289 & 0.0637 & 0.0329  & \textbf{0.2068} & 0.0371 \\
    L-FedAvg &  0.2870  & 0.1257 & 0.0636 & 0.0367  & 0.1979 & 0.0387 \\
    Krum Aggregation &  \textbf{0.3060} & \textbf{0.1411} & \textbf{0.0726} & 0.0377  & 0.2066 & 0.0395 \\
    \makecell[l]{Centralized \\ ViT B16+GPT2 \cite{islam2025visionlanguagemodelsautomatedchest}} & 0.2877 & 0.1273 & 0.0689 & \textbf{0.0435} & 0.2031 & \textbf{0.0403} \\
    \bottomrule
  \end{tabular}
\end{table*}

Federated Weighted Averaging (FedAvg) And L-FedAvg methods showed comparable performance. FedAvg slightly outperformed L-FedAvg in most metrics. FedAvg achieved a ROUGE-1 F1 score of 0.2928 and a BLEU score of 0.0371 whereas L-FedAvg scored 0.287 and 0.0387 for the same metrics respectively.

The Centralized ViT B16+GPT2 \cite{islam2025visionlanguagemodelsautomatedchest} approach demonstrated moderate performance. It had a ROUGE-4 F1 score of 0.0435 and BLEU score of 0.0403. Although it did not surpass Krum Aggregation in most metrics, it achieved the highest ROUGE-4 and BLEU scores among all approaches.

The results suggest that Krum Aggregation is the most robust method in terms of n-gram based Natural Language Generation(NLG) metrics. Centralized ViT B16+GPT2 approach provides competitive results particularly for higher-order ROUGE metrics. FedAvg and L-FedAvg perform consistently but fall behind in overall scores.
\begin{table}[ht]
  \caption{Evaluation with BERTScore and RaTEScore}
  \label{tab:bert_score_evaluation}
  \centering
  \begin{tabular}{lcccc}
    \toprule
    \textbf{Approach} & 
    \makecell{\textbf{BERTScore} \\ \textbf{Precision}} & 
    \makecell{\textbf{BERTScore} \\ \textbf{Recall}} & 
    \makecell{\textbf{BERTScore} \\ \textbf{F1 Score}} & 
    \textbf{RaTEScore} \\
    \midrule
    FedWeighted Average & \textbf{0.8509} & 0.8945 & 0.8720 & 60.93 \\
    L-FedAvg & 0.8437 & 0.8989 & 0.8703 & 61.99 \\
    Krum Aggregation & 0.8477 & \textbf{0.9003} & \textbf{0.8731} & \textbf{62.24} \\
    Centralized ViT B16+GPT2 \cite{islam2025visionlanguagemodelsautomatedchest} & 0.8392 & 0.9015 & 0.8691 & 53.47 \\
    \bottomrule
  \end{tabular}
\end{table}

\begin{figure}[htbp]
  \centering
  \begin{subfigure}[b]{0.40\linewidth}
      \includegraphics[width=\linewidth]{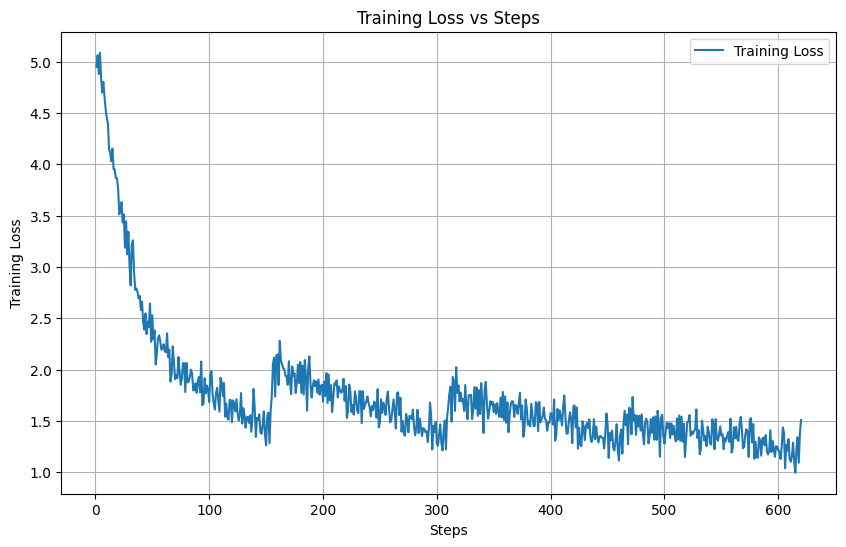}
      \caption{Client 1}
      \label{fig:ff_cl1_tr}
  \end{subfigure}
  \hfill
  \begin{subfigure}[b]{0.40\linewidth}
      \includegraphics[width=\linewidth]{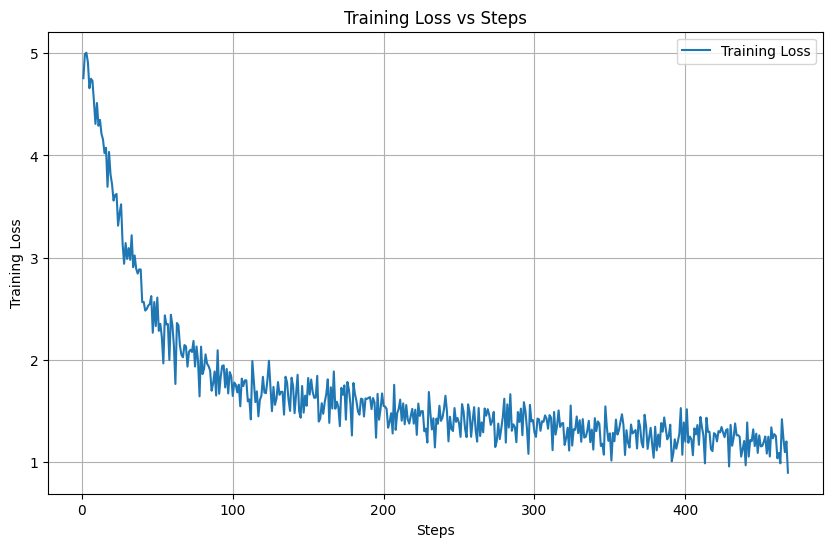}
      \caption{Client 2}
      \label{fig:ff_cl2_tr}
  \end{subfigure}

  \vspace{0.5em}

  \begin{subfigure}[b]{0.40\linewidth}
      \includegraphics[width=\linewidth]{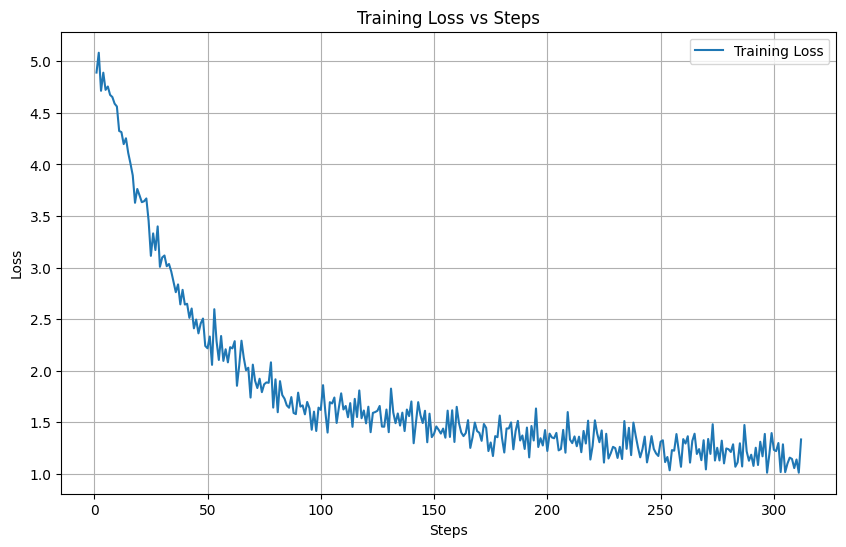}
      \caption{Client 3}
      \label{fig:ff_cl3_tr}
  \end{subfigure}
  \hfill
  \begin{subfigure}[b]{0.40\linewidth}
      \includegraphics[width=\linewidth]{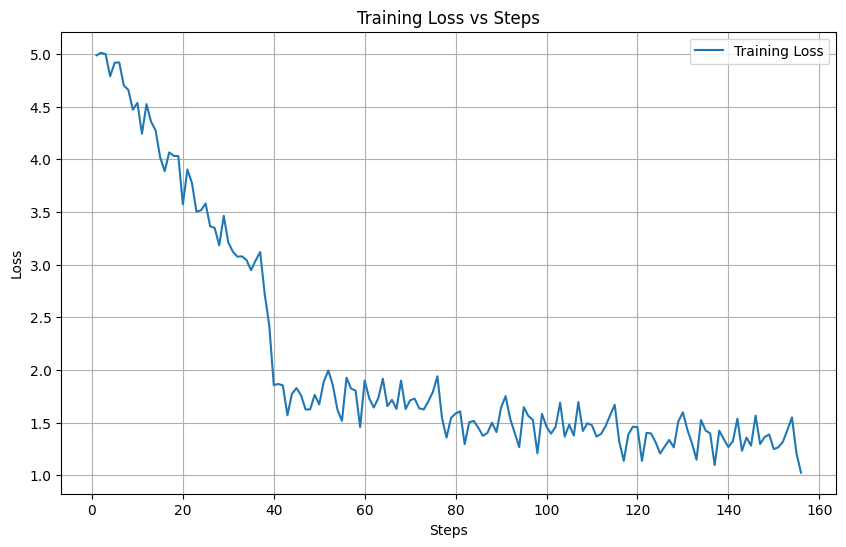}
      \caption{Client 4}
      \label{fig:ff_cl4_tr}
  \end{subfigure}

  \caption{Training Loss for Clients 1–4 in L-FedAvg}
  \label{fig:training_loss_clients}
\end{figure}

\vspace{1em} 

\begin{figure}[htbp]
  \centering
  \begin{subfigure}[b]{0.40\linewidth}
      \includegraphics[width=\linewidth]{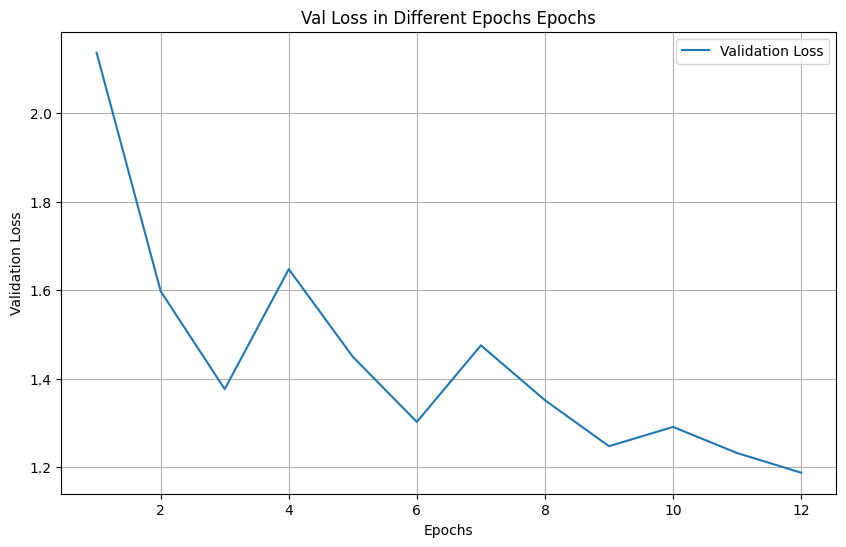}
      \caption{Client 1}
      \label{fig:ff_cl1_val}
  \end{subfigure}
  \hfill
  \begin{subfigure}[b]{0.40\linewidth}
      \includegraphics[width=\linewidth]{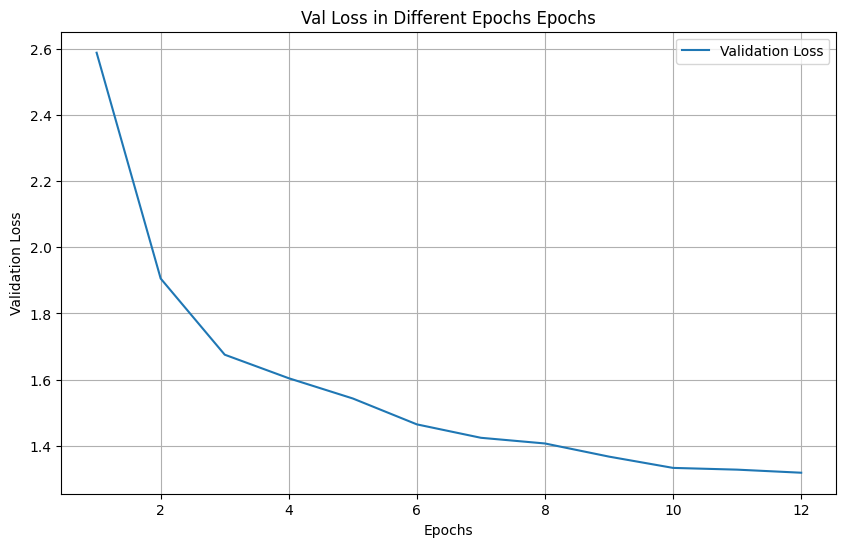}
      \caption{Client 2}
      \label{fig:ff_cl2_val}
  \end{subfigure}

  \vspace{0.5em}

  \begin{subfigure}[b]{0.40\linewidth}
      \includegraphics[width=\linewidth]{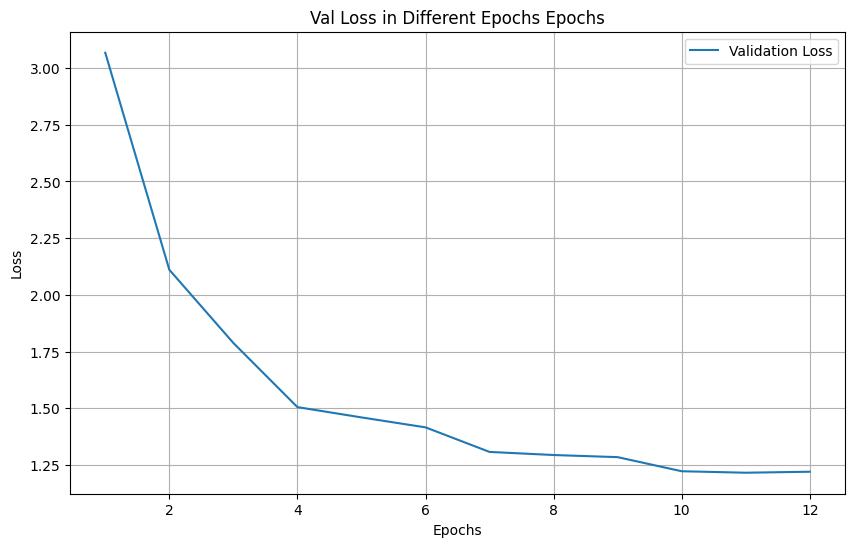}
      \caption{Client 3}
      \label{fig:ff_cl3_val}
  \end{subfigure}
  \hfill
  \begin{subfigure}[b]{0.40\linewidth}
      \includegraphics[width=\linewidth]{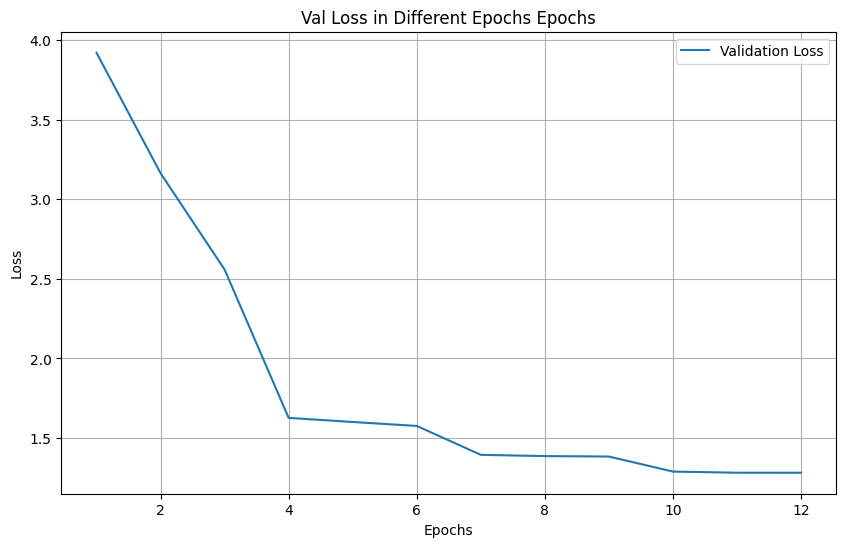}
      \caption{Client 4}
      \label{fig:ff_cl4_val}
  \end{subfigure}

  \caption{Validation Loss for Clients 1–4 in L-FedAvg}
  \label{fig:validation_loss_clients}
\end{figure}

We also evaluated the approaches using metrics such as BERTScore Precision, Recall and F1 to assess the semantic similarity between the generated reports and the ground truth. In Table \ref{tab:bert_score_evaluation}, Krum Aggregation achieved the highest BERTScore F1 of 0.8731. It also demonstrated balanced performance with a Precision of 0.8477 and a Recall of 0.9003, suggesting that this approach is robust in generating reports with both accuracy and coverage.

FedAvg and L-FedAvg methods showed comparable results. FedAvg achieved a slightly higher BERTScore F1 of 0.872 compared to 0.8703 for L-FedAvg. L-FedAvg obtained a better Recall at 0.8989, while Federated Weighted Averaging had slightly better Precision at 0.8509.

The Centralized ViT B16+GPT2 \cite{islam2025visionlanguagemodelsautomatedchest} approach achieved a BERTScore F1 of 0.8691. It also demonstrated the highest Recall among all methods at 0.9015. However, its Precision was the lowest among the approaches at 0.8392.

Again, Krum Aggregation emerged as the most effective method based on the BERTScore F1 metric. It demonstrated that this approach can be used to generate reports that are semantically accurate. The Centralized ViT B16+GPT2 approach demonstrated strong recall capabilities. FedAvg and L-FedAvg performed consistently well, with small trade-offs between precision and recall.

In terms of RaTEScore, which considers both semantic relevance and textual quality, Krum Aggregation again achieved the highest score (62.24), followed closely by L-FedAvg with 61.99. This indicates that L-FedAvg, while slightly trailing in BERTScore F1, excels in generating coherent and fluent reports and outperforms FedAvg and the centralized approach in terms of RaTEScore. This suggests that the loss-aware weighting strategy of L-FedAvg contributes to better overall text quality and relevance.

The training losses of different clients at various steps using the L-FedAvg approach are illustrated in Figure \ref{fig:training_loss_clients}. Specifically, Figure \ref{fig:ff_cl1_tr} shows the training loss for Client 1, Figure \ref{fig:ff_cl2_tr} for Client 2, Figure \ref{fig:ff_cl3_tr} for Client 3 and Figure \ref{fig:ff_cl4_tr} for Client 4. Client 1 had the most amount of data. The training loss consistently went downwards for client 1. However, some spikes can be observed after each federated round due to global model update. For client 2 and 3, the training loss decreased gradually with each step without any significant rise. The training loss for client 4 dropped massively after the first federated round.

The validation losses of the different clients across epochs using L-FedAvg approach are illustrated in Figure \ref{fig:validation_loss_clients}. Specifically, Figure \ref{fig:ff_cl1_val} shows the validation loss for Client 1, Figure \ref{fig:ff_cl2_val} for Client 2, Figure \ref{fig:ff_cl3_val} for Client 3 and Figure \ref{fig:ff_cl4_val} for Client 4. A similar pattern can be observed for the validation loss of client 1. After each federated round and global model update, the validation loss increased. For client 2 and 3, the validation loss consistently decreased. For client 4, the validation loss was heavily reduced after the first round which is consistent with the training loss.
\begin{figure}[htbp]
  \centering
  \begin{subfigure}[b]{0.4\linewidth}
    \includegraphics[width=\linewidth]{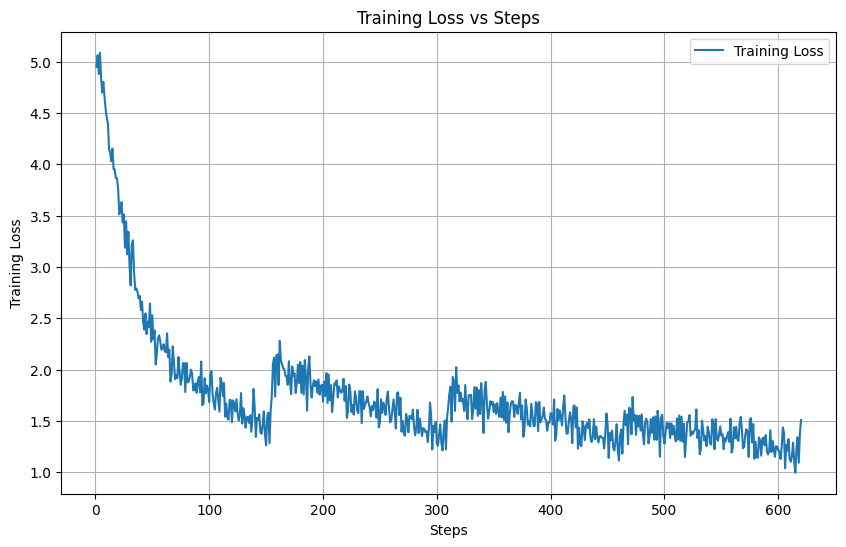}
    \caption{Client 1}
    \label{fig:k_cl1_tr}
  \end{subfigure}
  \hfill
  \begin{subfigure}[b]{0.4\linewidth}
    \includegraphics[width=\linewidth]{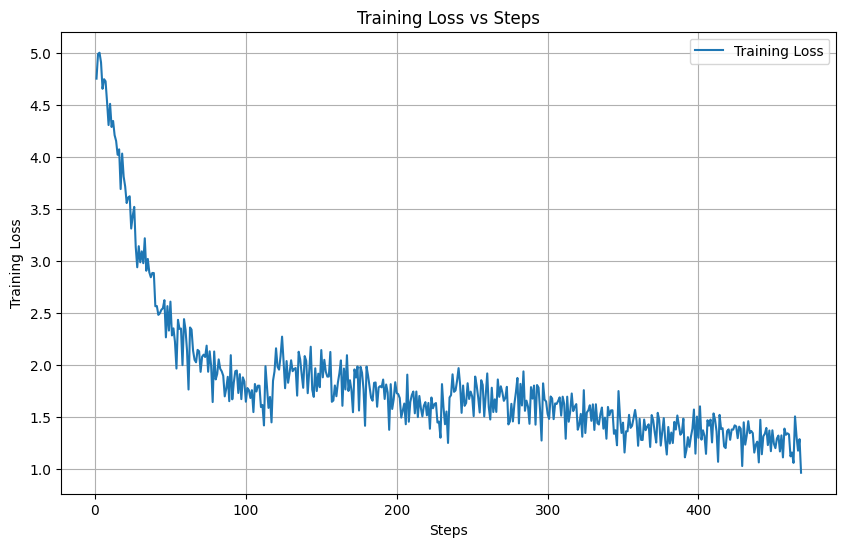}
    \caption{Client 2}
    \label{fig:k_cl2_tr}
  \end{subfigure}

  \vspace{0.5em}

  \begin{subfigure}[b]{0.4\linewidth}
    \includegraphics[width=\linewidth]{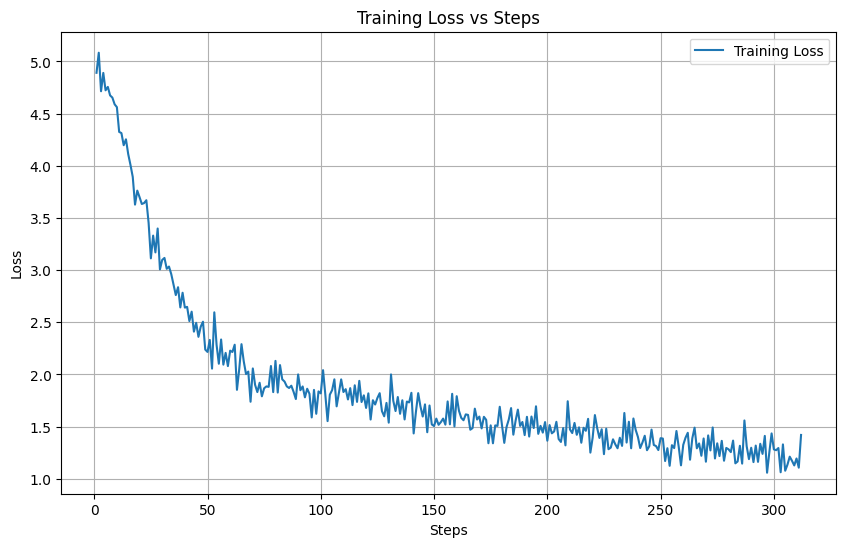}
    \caption{Client 3}
    \label{fig:k_cl3_tr}
  \end{subfigure}
  \hfill
  \begin{subfigure}[b]{0.4\linewidth}
    \includegraphics[width=\linewidth]{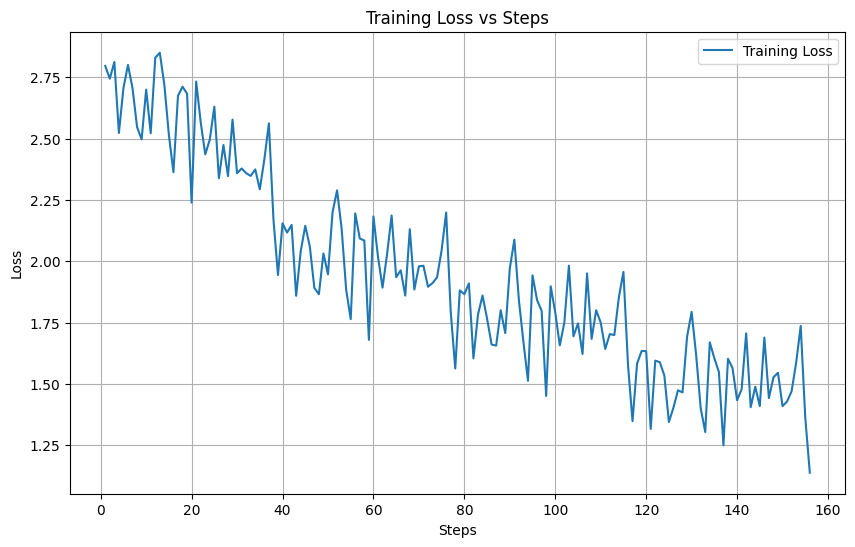}
    \caption{Client 4}
    \label{fig:k_cl4_tr}
  \end{subfigure}

  \caption{Training Loss for Clients 1–4 in Krum Aggregation}
  \label{fig:krum_training_loss_clients}
\end{figure}
\begin{figure}[htbp]
  \centering
  \begin{subfigure}[b]{0.40\linewidth}
    \includegraphics[width=\linewidth]{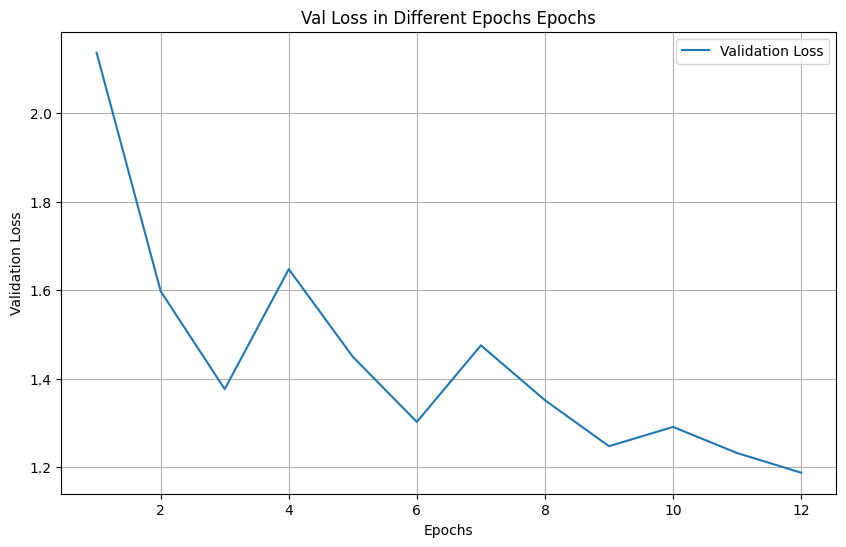}
    \caption{Client 1}
    \label{fig:k_cl1_val}
  \end{subfigure}
  \hfill
  \begin{subfigure}[b]{0.4\linewidth}
    \includegraphics[width=\linewidth]{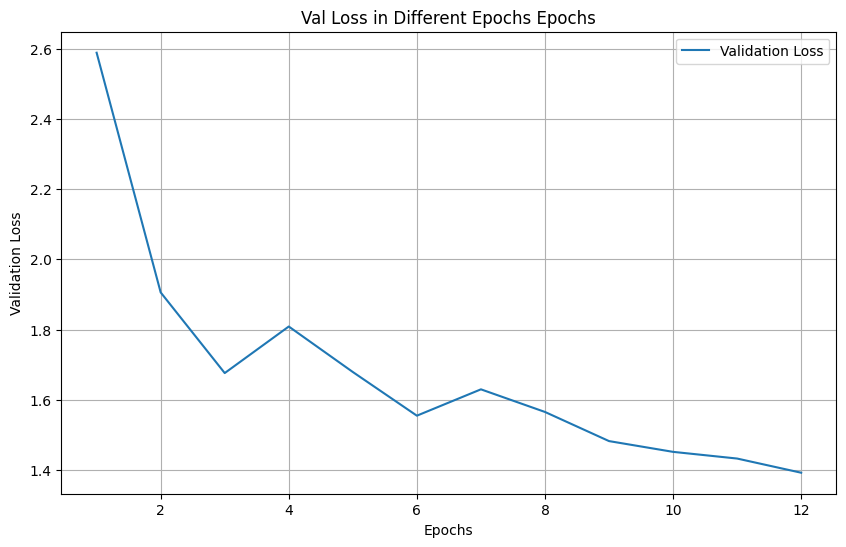}
    \caption{Client 2}
    \label{fig:k_cl2_val}
  \end{subfigure}

  \vspace{0.5em}

  \begin{subfigure}[b]{0.4\linewidth}
    \includegraphics[width=\linewidth]{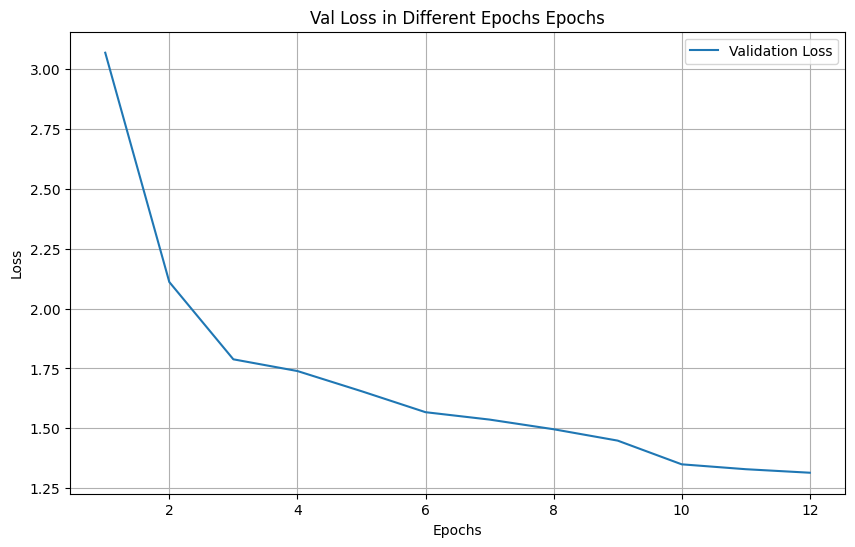}
    \caption{Client 3}
    \label{fig:k_cl3_val}
  \end{subfigure}
  \hfill
  \begin{subfigure}[b]{0.4\linewidth}
    \includegraphics[width=\linewidth]{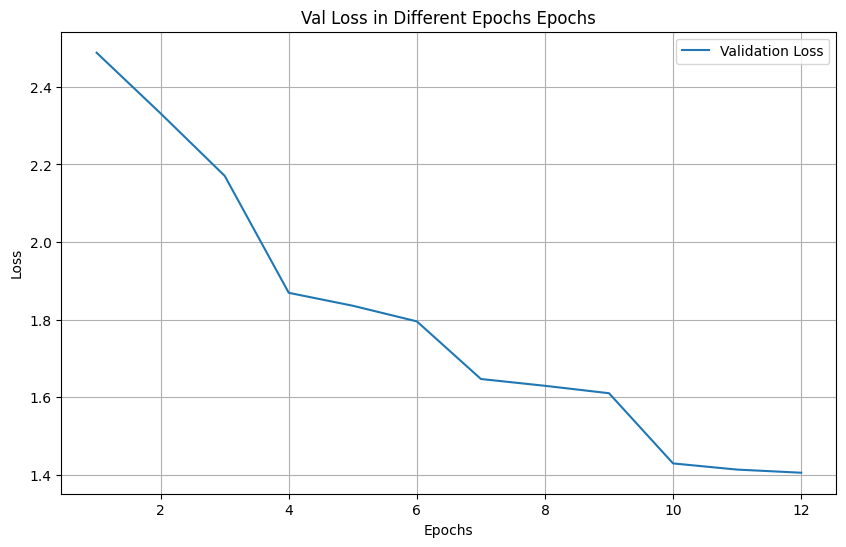}
    \caption{Client 4}
    \label{fig:k_cl4_val}
  \end{subfigure}

  \caption{Validation Loss for Clients 1–4 in Krum Aggregation}
  \label{fig:krum_validation_loss_clients}
\end{figure}
The training losses in different steps for all the clients using Krum Aggregation approach are illustrated in Figure \ref{fig:krum_training_loss_clients}. Specifically, Figure \ref{fig:k_cl1_tr} represents the training loss for Client 1, Figure \ref{fig:k_cl2_tr} for Client 2, Figure \ref{fig:k_cl3_tr} for Client 3 and Figure \ref{fig:k_cl4_tr} for Client 4. Again, we can see some spikes after each round in the training loss plot of client 1. The same pattern can be seen in client 2 training. The training loss for client 3 gradually decreased. However, since client 4 had the least amount of training data, a lot of oscillation can be observed in its training loss plot. 

The validation losses of different clients across epochs using the Krum aggregation approach are illustrated in Figure \ref{fig:krum_validation_loss_clients}. Specifically, Figure \ref{fig:k_cl1_val} shows the validation loss for Client 1, Figure \ref{fig:k_cl2_val} for Client 2, Figure \ref{fig:k_cl3_val} for Client 3 and Figure \ref{fig:k_cl4_val} for Client 4. We can notice the large spike in validation loss of client 1 after each federated round which is similar to its training loss. For client 2, we can see some smaller spikes in the validation loss after each round. For client 3, the validation loss consistently decreased. However, for client 4, an interesting pattern can be observed. The validation loss decreased massively in the first epoch of each federated round compared to other epochs of the round. So, the aggregation technique significantly improved this local client's performance.

\begin{table}[ht]
  \caption{Comparison with the existing literature}
  \label{tab:comparison_existing_literature}
  \centering
  \begin{tabular}{lcccc}
    \toprule
    \textbf{Method} & \textbf{BLEU} & \textbf{ROUGE-L F1} & \makecell{\textbf{BERTScore} \\ \textbf{F1}} & \textbf{RaTEScore} \\
    \midrule
    MAIRA-2 \cite{bannur2024maira} & 0.117 & 0.2740 & 0.5576 & -- \\
    CXRMate \cite{nicolson2024longitudinal} & 0.046 & 0.2820 & 0.3230 & -- \\
    EAST \cite{nicolson2024health} & 0.120 & 0.2651 & 0.5464 & -- \\
    XRaySwinGen \cite{magalhaes2024xrayswingen} & \textbf{0.124} & \textbf{0.3000} & -- & -- \\
    PromptMRG \cite{jin2024promptmrg} & 0.098 & 0.1600 & -- & -- \\
    NLGR-CCR \cite{liu2019clinically} & 0.102 & 0.2530 & -- & -- \\
    Centralized ViT B16+GPT2 \cite{islam2025visionlanguagemodelsautomatedchest} & 0.0403 & 0.2031 & 0.8691 & 53.47 \\
    FedViT-GPT2 (Krum, Ours) & 0.0426 & 0.2066 & \textbf{0.8731} & \textbf{62.24} \\
    \bottomrule
  \end{tabular}
\end{table}

\begin{figure}[ht]
  \centering
  \includegraphics[width=\linewidth]{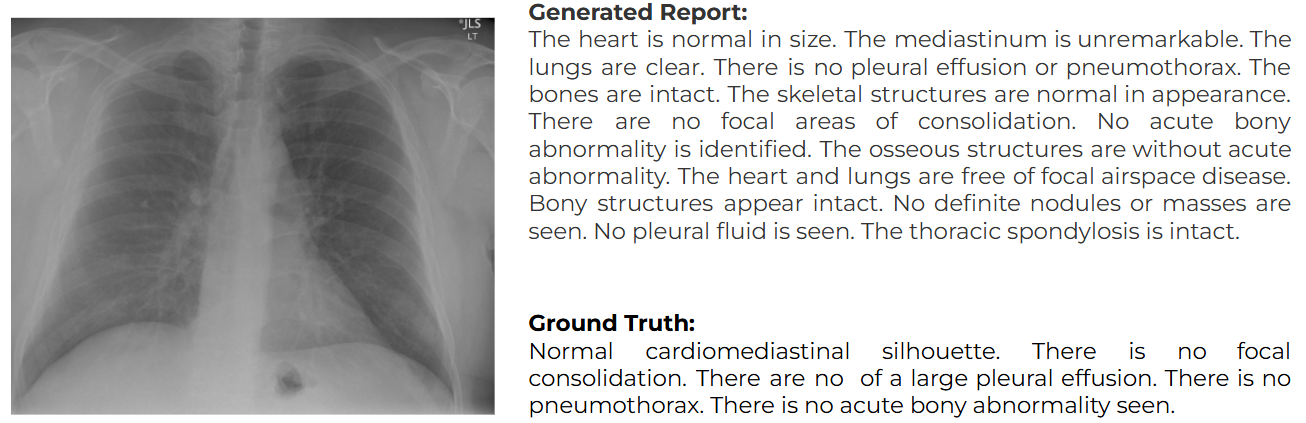}
  \caption{Generated report sample from our implementation with attached ground truth}
  \label{fig:generated_report1}
\end{figure}

Table \ref{tab:comparison_existing_literature} presents a comparative analysis of our proposed federated model (FedViT-GPT2 with Krum aggregation) against existing methods for chest X-ray report generation. While prior models such as XRaySwinGen \cite{magalhaes2024xrayswingen} and EAST \cite{nicolson2024health} have demonstrated competitive performance in terms of BLEU and ROUGE-L F1, they lack evaluation on deeper semantic metrics like BERTScore and RaTEScore. Our approach, although slightly behind XRaySwinGen in BLEU and ROUGE-L, significantly outperforms all baselines in semantic fidelity, achieving the highest BERTScore F1 (0.8731) and RaTEScore (62.24). These results highlight the strength of our model in generating semantically accurate and clinically coherent reports. Furthermore, our decentralized setup achieves comparable or better performance than centralized and fully supervised approaches such as Centralized ViT B16+GPT2 \cite{islam2025visionlanguagemodelsautomatedchest}, demonstrating the effectiveness of federated learning with robust aggregation in medical report generation.

Figure-\ref{fig:generated_report1} shows an example report generated through our implementation. From the attached ground truth, it is clearly visible that our model is capable of generating accurate and coherent report from a given x-ray image.

\section{Conclusion and Future Work}\label{sec6} 
In this paper we have evaluated different Federated Aggregation techniques for generating reports from chest x-ray images. Our experiment finds the best performance from the Krum Aggregation approach in the task of accurate and coherent report generation from input x-ray images. Due to limited number of data, we had to perform simulation with only four clients. Our approach can be easily extended for larger datasets and more clients. The issue of limited data should also be addressed in the future  to ensure reliable report generation.\\ \\ \\
\textbf{Conflict of interest}  The authors have no conflict of interest to declare relevant to this article’s content. Additionally, the authors have no relevant financial or non-financial interests to disclose. \\ \\ \\
\textbf{Data availability} Not applicable.
\backmatter

\bibliography{sn-bibliography}

\end{document}